\documentclass[aps,11pt,twoside,nofootinbib,tightenlines,superscriptaddress,preprintnumbers]{revtex4}

\usepackage{amsfonts,amssymb}
\usepackage{graphicx}
\usepackage{epic}

\newcommand{\<}{\langle}
\renewcommand{\>}{\rangle}

\newcommand{\be}{\begin{equation}}
\newcommand{\ee}{\end{equation}}
\newcommand{\bea}{\begin{eqnarray}}
\newcommand{\eea}{\end{eqnarray}}

\newcommand{\cond}[1]{\left\{\begin{array}{l@{~~~}l}#1\end{array}\right.}
\newcommand{\qph}[1]{quant-ph/#1}

\newcommand{\BPP}{{\mathrm{BPP}}}
\newcommand{\BQP}{{\mathrm{BQP}}}
\newcommand{\ent}{{\textsc{entrance}}}
\newcommand{\exit}{{\textsc{exit}}}
\renewcommand{\root}{{\textsc{root}}}

\renewcommand{\d}{{\mathrm{d}}}
\newcommand{\sech}{\mathop{\mathrm{sech}}\nolimits}
\newcommand{\Ai}{\mathop{\mathrm{Ai}}\nolimits}
\newcommand{\poly}{\mathop{\mathrm{poly}}\nolimits}
\newcommand{\col}{\mathop{\mathrm{col}}\nolimits}

\newcommand\symProb{\mathop{\mathrm{Pr}}\displaylimits}
\newcommand\symExpec{\mathop{\mathrm{E}}\displaylimits}
\def\prob#1#2{\symProb_{#1}\left[ #2 \right]}
\def\expec#1#2{\symExpec_{#1}\left[ #2 \right]}

\newtheorem{theorem}{Theorem}
\newtheorem{lemma}[theorem]{Lemma}
\newtheorem{game}{Game}

\newcommand{\qed}{\rule{7pt}{7pt}}
\newenvironment{proof}
  {\trivlist\item\noindent{\bf Proof}~}
  {\qed\endtrivlist}

\begin{document}

\title{Exponential algorithmic speedup by quantum walk}

\author{Andrew M. Childs}
\email[]{amchilds@mit.edu}
\affiliation{Center for Theoretical Physics,
             Massachusetts Institute of Technology,
             Cambridge, MA 02139, USA}

\author{Richard Cleve}
\email[]{cleve@cpsc.ucalgary.ca}
\affiliation{Department of Computer Science,
             University of Calgary,
             Calgary, Alberta, Canada T2N 1N4}

\author{Enrico Deotto}
\email[]{deotto@mitlns.mit.edu}
\affiliation{Center for Theoretical Physics,
             Massachusetts Institute of Technology,
             Cambridge, MA 02139, USA}

\author{Edward Farhi}
\email[]{farhi@mit.edu}
\affiliation{Center for Theoretical Physics,
             Massachusetts Institute of Technology,
             Cambridge, MA 02139, USA}

\author{Sam Gutmann}
\email[]{sgutm@neu.edu}
\affiliation{Department of Mathematics,
             Northeastern University,
             Boston, MA 02115, USA}

\author{Daniel A. Spielman}
\email[]{spielman@math.mit.edu}
\affiliation{Department of Mathematics,
             Massachusetts Institute of Technology,
             Cambridge, MA 02139, USA}

\date[]{25 October 2002}

\preprint{MIT-CTP \#3309}

%%%%%%%%%%%%%%%%%%%%%%%%%%%%%%%%%%%%%%%%%%%%%%%%%%%%%%%%%%%%%%%%%%%%%%%%%%%%%%%
% Abstract

\begin{abstract}
We construct an oracular (i.e., black box) problem that can be solved
exponentially faster on a quantum computer than on a classical computer.
The quantum algorithm is based on a continuous time quantum walk, and thus
employs a different technique from previous quantum algorithms based on
quantum Fourier transforms.  We show how to implement the quantum walk
efficiently in our oracular setting. We then show how this quantum walk
can be used to solve our problem by rapidly traversing a graph.  Finally,
we prove that no classical algorithm can solve this problem with high
probability in subexponential time.
\end{abstract}

\maketitle

%%%%%%%%%%%%%%%%%%%%%%%%%%%%%%%%%%%%%%%%%%%%%%%%%%%%%%%%%%%%%%%%%%%%%%%%%%%%%%%
\section{Introduction} \label{sec:intro}

A primary goal of the field of quantum computation is to determine when
quantum computers can solve problems faster than classical computers.
Exponential quantum speedup has been demonstrated for a number of
different problems, but in each case, the quantum algorithm for solving
the problem relies on the quantum Fourier transform.  The purpose of this
paper is to demonstrate that exponential speedup can be achieved by a
different algorithmic technique, the quantum walk.  We show that quantum
walks can solve an oracular computational problem exponentially
faster than any classical algorithm.

Oracular problems provided the first examples of algorithmic speedup using a
quantum instead of a classical computer.  An oracular problem is specified in
terms of a black box function, and the goal is to find some property of
that function using as few queries to the black box as possible.  In this
setting, Deutsch gave an example of a problem that can be solved on a
quantum computer using one query, but that requires two queries on a
classical computer \cite{Deu85}.  Deutsch and Josza generalized this
problem to one that can be solved exactly on a quantum computer in
polynomial time, but for which an exact solution on a classical computer
requires exponential time \cite{DJ92}.  However, this problem can be
solved with high probability in polynomial time using a probabilistic
classical algorithm.  Bernstein and Vazirani gave the first example of a
superpolynomial separation between probabilistic classical and quantum
computation \cite{BV93}, and Simon gave another example in which the
separation is exponential \cite{Sim94}.

Quantum computers can also provide computational speedup over the best
known classical algorithms for non-oracular problems.  The first such
example was provided by Shor, who gave polynomial-time algorithms for
integer factorization and the discrete logarithm \cite{Sho94}.  A number
of generalizations and variations of these algorithms have been discovered
for solving both oracular and non-oracular problems with exponential
speedup (e.g.,
\cite{Kit95,ME99,BCW00,DH00,HRT00,DHI01,GSVV01,IMS01,Wat01,Hal02}).
All of them are fundamentally based on quantum Fourier transforms.

We would like to find new techniques for achieving algorithmic speedup
with quantum computers.  Since many classical algorithms are based on
random walks, it is natural to ask whether a quantum analogue of a random
walk process might be useful for quantum computation.  This idea was
explored by Farhi and Gutmann \cite{FG98}, who proposed the model of a
quantum walk used in the present paper.\footnote{The term {\em quantum
random walk} is somewhat misleading since the dynamics of this process are
coherent, so we use the term {\em quantum walk} in this paper.}  They gave
an example of a graph in which the time to propagate between two vertices
(i.e., the hitting time) is exponentially faster for the quantum walk than
for the corresponding classical random walk.  A simpler example of a graph
with an exponentially faster quantum hitting time was given in
\cite{CFG02}.  However, as we explain in Section \ref{sec:problem}, these
results do not imply algorithmic speedup.  In the present paper, we modify
the example presented in \cite{CFG02} to construct an oracular problem
that can be solved efficiently using a quantum walk, but that no classical
algorithm can solve in subexponential time.  Although the quantum walk is
defined as a continuous time quantum process, we show how to implement it
in the conventional quantum computing paradigm with the structure of the
graph provided in the form of an oracle.  In other words, the algorithm
consists of a polynomial-length sequence of applications of the oracle and
unitary gates that act on a few qubits at a time.

We note that the quantum analogue of a classical random walk is not
unique, and other models have been proposed.  Whereas the states of the
continuous time quantum walk used in this paper lie in a Hilbert space
spanned by states corresponding to the vertices in the graph, these
alternative models operate in discrete time and make use of an extra state
space (e.g., to implement a ``quantum coin'')
\cite{ADZ93,Mey96,Wat01b,AAKV01,NV01}.  As far as we know, our results are
the first example of algorithmic speedup based on either discrete or
continuous time quantum walks.

The structure of this paper is as follows.  In Section \ref{sec:problem},
we construct a family of graphs and use it to define our problem.  In
Section \ref{sec:algorithm}, we present a quantum algorithm for solving
this problem in polynomial time with high probability, and in Section
\ref{sec:lowerbound}, we show that the problem cannot be solved
classically in subexponential time with high probability.  We conclude
with a discussion of the results in Section \ref{sec:discussion}.

%%%%%%%%%%%%%%%%%%%%%%%%%%%%%%%%%%%%%%%%%%%%%%%%%%%%%%%%%%%%%%%%%%%%%%%%%%%%%%%
\section{The problem}\label{sec:problem}

In this section we describe in detail the problem we address in this
paper.  The problem involves determining a certain property of a graph
whose structure is provided in the form of an oracle.  A graph $G$ is a
set of $N$ vertices and a set of edges that specify which pairs of
vertices are connected.

\begin{figure}
\includegraphics[width=280pt]{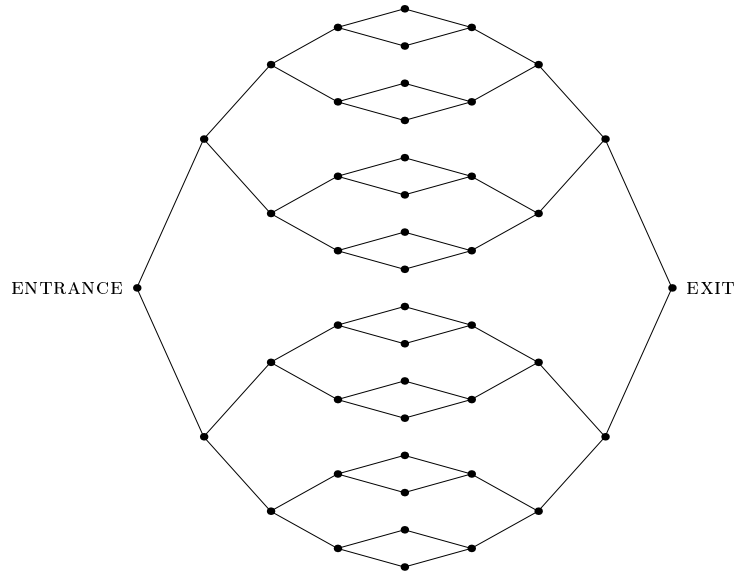}
\caption{The graph $G_4$.}
\label{fig:graph}
\end{figure}

Our problem is based on a generalization of the graphs presented in
\cite{CFG02}.  In that paper, the authors consider a sequence of graphs
$G_n$ consisting of two balanced binary trees of height $n$ with the $2^n$
leaves of the left tree identified with the $2^n$ leaves of the right tree
in the simple way shown in Figure \ref{fig:graph} (for $n=4$).  A
classical random walk starting at the left root of this graph requires
exponentially many steps (in $n$) to reach the right root.  However, the
continuous time quantum walk traverses $G_n$ in a time linear in $n$.

\begin{figure}
\includegraphics[width=284pt]{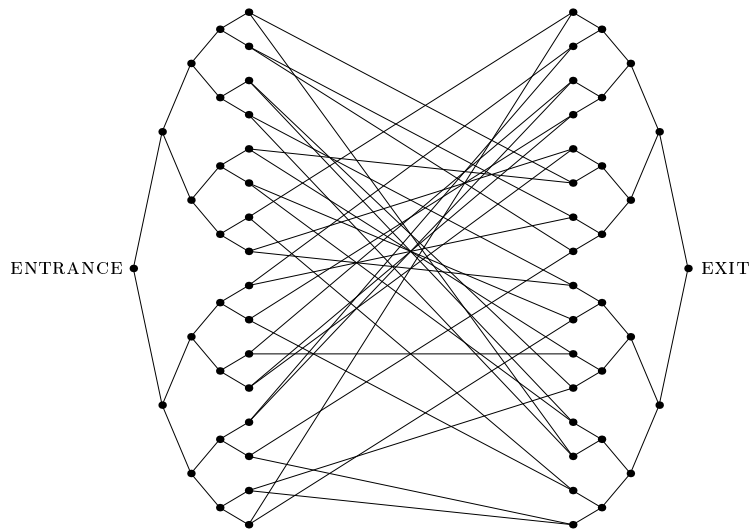}
\caption{A typical graph $G^{\prime}_4$.}
\label{fig:graphprime}
\end{figure}

In this paper, we modify the graphs in the previous example so that the
quantum walk is exponentially better not only than the corresponding
classical random walk, but also than {\em any} classical algorithm one can
design to traverse the graph.
Before we describe the modified graphs, we provide an oracular (black box)
setting in which graphs can be specified, and where our notion of an
algorithm traversing a graph can be made precise.  The vertices are
named with randomly chosen $2n$-bit strings, and the oracle only refers to
the vertices by their names.  Because $G_n$ contains $O(2^n)$ vertices,
$n+O(1)$ bits would be sufficient to give each vertex a unique name, but
we use $2n$ bits for each name so that there are exponentially more
possible names than vertices in the graph.  The oracle takes a $2n$-bit
string as input, and if that name corresponds to an actual vertex in the
graph, the oracle outputs the names of the adjacent vertices.  There are
also two identifiable vertices called $\ent$ and $\exit$, which in the
case of $G_n$ are the left root and the right root of the trees,
respectively.  The traversal problem is:
\begin{itemize}
\item Given an oracle for the graph and the name of the $\ent$, find the
name of the $\exit$.
\end{itemize}

However, even with $G_n$ given by such an oracle, the $\exit$ can be found
in polynomial time using a classical algorithm that is {\em not} a random
walk.  The key is that we can always tell whether a particular vertex is
in the central column by checking its degree.  We begin at the $\ent$.  At
each vertex, we query the oracle and move to one of the two unvisited
adjacent vertices.  After $n$ steps, we reach the central column and then
proceed moving to unvisited adjacent vertices in the right tree, checking
the degree at each step.  If we discover that after $k$ steps we are again
at a central vertex, we know that the wrong move happened after $k/2$
steps ($k$ can only be even due to the structure of $G_n$) and we can
backtrack to that point and take the other edge.  After only $O(n^2)$
steps this procedure will reach the $\exit$.\footnote{As an aside, we note
that there is also a polynomial-time classical traversal algorithm for the
$n$-dimensional hypercube (where one vertex is designated as the $\ent$
and the $\exit$ is the unique vertex whose distance from $\ent$ is $n$).
For a description of the algorithm, see Appendix \ref{app:hypercube}.
This means that there can be no exponential algorithmic speedup using
quantum walks to traverse the hypercube, even though both continuous and
discrete time quantum walks have been shown to reach the $\exit$ in a
polynomial number of steps in a non-oracular setting \cite{MR01,Kem02}.}

We now describe how the graphs $G_n$ are modified so that they cannot be
traversed efficiently with a classical algorithm.  We choose a graph
$G_n'$ at random from a particular distribution on graphs.  A typical
graph $G_n'$ is shown in Figure \ref{fig:graphprime} (for $n=4$).  The
distribution is defined as follows.  The graph again consists of two
balanced binary trees of height $n$, but instead of identifying the
leaves, they are connected by a random cycle that alternates between the
leaves of the two trees.  In other words, we choose a leaf on the left at
random and connect it to a leaf on the right chosen at random.  Then we
connect the latter to a leaf on the left chosen randomly among the
remaining ones.  We continue with this procedure, alternating sides, until
every leaf on the left is connected to two leaves on the right (and vice
versa).  As we will show in Section \ref{sec:lowerbound}, no
subexponential time classical algorithm can with non-negligible
probability find the $\exit$ given the $\ent$ and a typical oracle for
$G_n'$.

We will show in Sections \ref{subsec:line} and \ref{subsec:hitting} that
the quantum walk goes from {\scshape entrance} to {\scshape exit} in any
graph $G_n'$ in polynomial time.  However, to construct a quantum
algorithm, we must show how to {\it implement} the quantum walk, i.e., how
to efficiently simulate the appropriate Hamiltonian evolution on a quantum
computer.  We do this in the conventional paradigm for oracular quantum
computation, and our implementation will work for a general graph.  The
algorithm will consist of a sequence of unitary operators that are either
the oracle or act on a few qubits at a time.  We address the general
implementation issue in Section \ref{subsec:implement}.  In order to
implement the quantum walk on a general graph, we must also be given a
consistent coloring of the edges of the graph.  In other words, there must
be $k=\poly(\log(\# \mathrm{vertices}))$ colors assigned to the edges of
the graph such that all the edges incident on a given vertex are different
colors.  The input to the oracle consists of both the name of a vertex and
a particular color, and the output name will depend on the input color, as
explained in detail in Section \ref{subsec:implement}.

In principle, knowledge of such a coloring might help a classical
algorithm solve our problem, because there could be correlations between
the colors of the edges in different places in the graph.  However, we
will show in Section \ref{sec:lowerbound} that there exists a consistent
(random) coloring of any graph $G_n'$, using only nine colors (independent
of $n$), that is completely useless to any classical algorithm.
Consequently, this modification of the oracle can be made irrelevant from
the classical point of view.  For an alternative implementation of the
quantum walk on the graphs $G_n'$ that does not require colors, see
Appendix \ref{app:bipartite}.

%%%%%%%%%%%%%%%%%%%%%%%%%%%%%%%%%%%%%%%%%%%%%%%%%%%%%%%%%%%%%%%%%%%%%%%%%%%%%%%
\section{Quantum algorithm}\label{sec:algorithm}

In this section, we discuss quantum walks in general and then use them to
present a quantum algorithm for solving our problem.  We begin by
describing the model of a continuous time quantum walk on any graph in
Section \ref{subsec:qwalk}.  We define the quantum walk to be
Schr\"odinger evolution with the Hamiltonian equal to (a constant multiple
of) the adjacency matrix of the graph.  We explain how to implement the
quantum walk on a general graph in an oracular setting on a quantum
computer in Section \ref{subsec:implement}.  In Section \ref{subsec:line}
we show that the quantum walk on the graph $G_n'$ starting at the $\ent$
can be viewed as propagation on a discrete line, and we give a sequence of
arguments explaining why the walk propagates through the graph in linear
time.  Finally, in Section \ref{subsec:hitting}, we prove that the quantum
walk reaches the $\exit$ using a polynomial number of calls to the oracle.

%%%%%%%%%%%%%%%%%%%%%%%%%%%%%%%%%%%%%%%%%%%%%%%%%%%%%%%%%%%%%%%%%%%%%%%%%%%%%%%
\subsection{Quantum walk}\label{subsec:qwalk}

We now formally describe our model of a quantum walk on a general graph
$G$.  We write $a \in G$ to denote that the vertex $a$ is in the graph and
$aa' \in G$ to denote that the edge joining vertices $a$ and $a'$ is in
the graph.  We let $d(a)$ denote the degree of vertex $a$, i.e., the
number of edges incident on that vertex.

Our model of a quantum walk is defined in close analogy to a continuous
time classical random walk, which is a Markov process.  In a continuous
time classical random walk, there is a fixed probability per unit time
$\gamma$ of moving to an adjacent vertex.  In other words, from any
vertex, the probability of jumping to any connected vertex in a time
$\epsilon$ is $\gamma \epsilon$ (in the limit $\epsilon \to 0$).  If the
graph has $N$ vertices, this classical random walk can be described by the
$N \times N$ infinitesimal generator matrix $K$ defined by
\be
  K_{aa'} = \cond{ 
      \gamma       & a \ne a', \ aa' \in G \\
      0            & a \ne a', \ aa' \notin G \\
      -d(a) \gamma & a=a' \,.}
\ee
If $p_a(t)$ is the probability of being at vertex $a$ at time $t$, then
\be
  {\d p_a(t) \over \d t} = \sum_{a'} K_{aa'} \, p_{a'}(t)
\,.
\label{eq:diffeq}
\ee
Note that because the columns of $K$ sum to zero, an initially normalized
distribution remains normalized, i.e., $\sum_a p_a(t) = 1$ for all $t$.

Now consider quantum evolution in an $N$-dimensional Hilbert space spanned
by states $|1\>, |2\>, \ldots, |N\>$ corresponding to the vertices of $G$.
If the Hamiltonian is $H$, then the dynamics of the system are determined
by the Schr\"odinger equation,
\be
  i {{\mathrm d} \over {\mathrm d}t} \<a|\psi(t)\> 
    = \sum_{a'} \<a|H|a'\> \<a'|\psi(t)\>
\,.
\label{eq:schrodinger}
\ee
Note the similarity between (\ref{eq:diffeq}) and (\ref{eq:schrodinger}).
A natural quantum analogue to the continuous time classical random walk
described above is given by the Hamiltonian with matrix
elements~\cite{FG98}
\be
  \<a|H|a'\> = K_{aa'}
\,.
\label{eq:graphham1}
\ee
In the quantum case, there is no need for the columns of $H$ to sum to zero;
we only require $H$ to be Hermitian. We set the diagonal to zero
for simplicity.  In this paper, we define a quantum walk in terms of a
Hamiltonian with matrix elements
\be
  \<a|H|a'\> = \cond{ \gamma & a \ne a', \ aa' \in G \\
                      0      & \textrm{otherwise.} }
\label{eq:graphham}
\ee
In other words, $H$ is the adjacency matrix of the graph times $\gamma$.
Because $H$ is Hermitian, probability is conserved, i.e., $\sum_a
|\<a|\psi(t)\>|^2=1$ for all $t$.

%%%%%%%%%%%%%%%%%%%%%%%%%%%%%%%%%%%%%%%%%%%%%%%%%%%%%%%%%%%%%%%%%%%%%%%%%%%%%%%
\subsection{Implementing the quantum walk}\label{subsec:implement}

In this section, we describe how to implement the quantum walk on a
general graph $G$ using a universal quantum computer.  Our goal is to
simulate the unitary evolution $e^{-i H t}$ with $H$ given by
(\ref{eq:graphham}).  We want to do this with the graph given to us in the
form of an oracle.  However, in the oracular setting, we have only been
able to implement the quantum walk on a general graph using additional
structure: we require that the graph comes with a consistent coloring.  We
assume that each edge of the graph is assigned a color, where the total
number of colors is $k=\poly(\log N)$.  We say that a coloring is
consistent if no vertex is incident with two edges of the same color.
Although we might want to use more than the minimal number of colors for a
particular application, any graph whose maximum degree is $\Delta$ can be
consistently colored with at most $\Delta+1$ colors \cite{Viz64}.  For the
{\it specific} problem addressed in this paper, we will show that the
oracle can provide a consistent coloring that cannot be used by any
classical algorithm to help solve the problem, since a classical algorithm
could make up the coloring as it goes.  Alternatively, we could implement
the quantum walk for this particular problem using a similar idea in which
the coloring is generated by the quantum algorithm.  We outline this
construction in Appendix \ref{app:bipartite}.

We now give our method for implementing the quantum walk on a general
graph $G$ for which the oracle has assumed a particular consistent
coloring of the graph.  Let $n=\lceil \log N \rceil$ so it takes $n$ bits
to list the vertices of the graph $G$.  In the classical setting, the
black box takes two inputs, a name $a$ given as a $2n$-bit string and a
color $c$.  If the input name $a$ corresponds to a vertex that is incident
with an edge of color $c$, then the output is the name of the vertex
joined by that edge.  If $a$ is not the name of a vertex or if $a$ is the
name of a vertex but there is no incident edge of color $c$, then the
output is the special bit string $11\ldots1$, which is not the name of any
vertex.  For shorthand, we write $v_c(a)=a'$, where $a$ is the input name,
$c$ is the color, and $a'$ is the output name.  If $a' \ne 11\ldots1$,
then $a'$ is the name of the vertex that is connected by an edge of color
$c$ to the vertex named $a$.  Note that $v_c(v_c(a))=a$ for $v_c(a) \ne
11\ldots1$, which is a crucial ingredient in our implementation of the
quantum walk.

The unitary quantum black box corresponding to this classical black box is
described as follows.  Let $a,b$ be $2n$-bit strings and let $c$ be a
color.  Then the action of the quantum black box $U$ associated with the
graph $G$ is
\be
  U |a,b,c\> = |a,b \oplus v_c(a),c\>
\label{eq:graphoracle}
\ee
where $\oplus$ denotes bitwise addition modulo 2.  However, we will never
need to query $U$ with a superposition of different colors, so for our
purposes, it is sufficient to omit the color register and assume that we
have access to the $k$ unitary transformations
\be
  U_c|a,b\> = |a,b \oplus v_c(a)\>
\,,
\ee
one for each of the $k$ possible colors.  Clearly, this operation can be
performed using the oracle (\ref{eq:graphoracle}).  In fact, by checking
whether $v_c(a)=11\ldots1$, it is also straightforward to extend the
Hilbert space by a single qubit and perform each of the $k$
transformations
\be
  V_c|a,b,r\> = |a,b \oplus v_c(a),r \oplus f_c(a)\>
\,,
\label{eq:simpleoracle}
\ee
where
\be
  f_c(a) = \cond{
      0 & v_c(a) \ne 11\ldots1 \\
      1 & v_c(a) = 11\ldots1 \,,}
\ee
using only a few calls to (\ref{eq:graphoracle}).\footnote{Note that in
(\ref{eq:simpleoracle}), the register containing $r$ consists of a single
qubit, as opposed to (\ref{eq:graphoracle}), in which $c$ represents one
of $k$ colors.}
In summary, for our purposes, we may view the oracle as giving us the
ability to perform each of the $k$ unitary transformations $V_c$ given by
(\ref{eq:simpleoracle}), where $a,b$ are $2n$-bit strings and $r$ is a
single bit.

The Hilbert space that $V_c$ acts on in (\ref{eq:simpleoracle}) has
dimension $2^{4n+1}$, which is much larger than the number of vertices in
the graph.  We want the quantum evolution $e^{-i H t}$ to take place in an
$N$-dimensional subspace.  To this end, we will show how to implement the
quantum walk given the oracles $V_c$ in the sense that we will construct a
Hamiltonian $H$ satisfying
\be
  H |a, 0, 0\> = \sum_{c:\, v_c(a) \in G} |v_c(a),0,0\>
\,.
\label{eq:ham}
\ee
This means that if we start our evolution in the subspace of states of the
form $|a,0,0\>$ where $a \in G$, then we will remain in this
$N$-dimensional subspace.

In a non-oracular setting, we say we can efficiently simulate any
$m$-qubit Hamiltonian $H$ if we can approximate the unitary evolution
$e^{-i H t}$ for any $t=\poly(m)$ using a polynomial number of one- and
two-qubit unitary gates.  In our oracular setting, we regard a simulation
of $H$ as efficient if it can approximate $e^{-i H t}$ for any
$t=\poly(m)$ using a polynomial number of calls to the oracle and a
polynomial number of additional one- and two-qubit gates.  In general, in
either setting, it is not easy to answer the question of whether a
particular Hamiltonian can be simulated efficiently.\footnote{Certainly
{\em most} Hamiltonians cannot be simulated efficiently, since there are
many more possible $m$-qubit Hamiltonians than strings of $\poly(m)$ one-
and two-qubit gates.} However, there are some useful standard tools for
simulating Hamiltonians.  The following is a list of five such tools, four
of which will be used in our construction.

\begin{enumerate}
\item {\em Local terms.}  If $H$ acts on $O(1)$ qubits, it can be
simulated.  This is simply because any operator acting on a constant
number of qubits can be approximated using a constant number of one- and
two-qubit gates.
\label{sim:local}

\item {\em Linear combination.}  If we can simulate $H_1, \ldots H_k$,
then we can simulate $H_1 + \cdots + H_k$ as a result of the Lie product
formula
\be
  e^{-i(H_1+\cdots+H_k)t}
    = \left(e^{-i H_1 t/j} \cdots e^{-i H_k t/j}\right)^j 
      + O(k \|[H_p,H_q]\| t^2 / j)
\,,
\label{eq:lie}
\ee
where $\|[H_p,H_q]\|$ is the largest norm of a commutator of two of the
Hamiltonians.  This is a powerful tool for the simulation of {\em
physical} quantum systems, which can typically be expressed as a sum of
local terms \cite{Llo96}.
\label{sim:linear}

\item {\em Commutation.}  If we can simulate $H_1$ and $H_2$, then we can
simulate $i[H_1,H_2]$.  This is a consequence of the identity
\be
  e^{[H_1,H_2]t} = \lim_{j \to \infty} 
    \left(e^{-i H_1 t/\sqrt j} e^{-i H_2 t/\sqrt j} 
          e^{ i H_1 t/\sqrt j} e^{ i H_2 t/\sqrt j}\right)^j
\,.
\ee
Using linear combination and commutation, it is possible to simulate any
Hamiltonian in the Lie algebra generated by a set of Hamiltonians.  We
will not need to use commutation in the present paper, but we include it
for completeness.
\label{sim:comm}

\item {\em Unitary conjugation.}  If we can simulate $H$ and we can
efficiently perform the unitary operation $U$, then we can simulate $U H
U^\dag$.  This follows from the simple fact that $U e^{-i H t} U^\dag =
e^{-i U H U^\dag t}$.
\label{sim:unitary}

\item {\em Tensor product.}  If $H_j$ each act on $O(1)$ qubits and the
product of their eigenvalues is efficiently computable, then we can
simulate $\bigotimes_j H_j$.  This can be done with a generalization of a
technique discussed in Section 4.7.3 of \cite{NC00}.  We will not present
the general construction, but we show its application to a specific
Hamiltonian below.  Note that $\bigotimes_j H_j$ may be highly nonlocal,
so it may look very different from the Hamiltonian of a physical system.
\label{sim:tensor}
\end{enumerate}

We now use these tools to construct the quantum walk on a graph $G$
specified by the oracles $V_c$ given by (\ref{eq:simpleoracle}).  In other
words, we show how to efficiently simulate (\ref{eq:ham}).  The 
Hilbert space will consist of states of the form $|a, b, r\>$, where $a$
and $b$ are $2n$-bit strings and $r$ is a single bit.  The states that
correspond to vertices are $|a, 0, 0\>$, where $a$ is the name of a vertex
in the graph.

We begin by showing how to simulate the evolution generated by the
Hermitian operator $T$ satisfying
\bea
  T |a, b, 0\> &=& |b, a, 0\> \\
  T |a, b, 1\> &=& 0
\,.
\eea
We will use $T$ as a building block in our simulation of $H$.  Our
simulation of $T$ uses only $O(n)$ one- and two-qubit gates.

\begin{figure}
\setlength{\unitlength}{2pt}
\begin{picture}(180,105)
\put(10,5){\line(1,0){78}}
\put(102,5){\line(1,0){78}}
\put(10,15){\line(1,0){84}}
\put(96,15){\line(1,0){84}}
\put(10,25){\line(1,0){10}}
\put(40,25){\line(1,0){39}}
\put(81,25){\line(1,0){28}}
\put(111,25){\line(1,0){39}}
\put(170,25){\line(1,0){10}}
\put(10,35){\line(1,0){10}}
\put(40,35){\line(1,0){110}}
\put(170,35){\line(1,0){10}}
\put(10,65){\line(1,0){10}}
\put(40,65){\line(1,0){19}}
\put(61,65){\line(1,0){68}}
\put(131,65){\line(1,0){19}}
\put(170,65){\line(1,0){10}}
\put(10,75){\line(1,0){10}}
\put(40,75){\line(1,0){110}}
\put(170,75){\line(1,0){10}}
\put(10,90){\line(1,0){10}}
\put(40,90){\line(1,0){9}}
\put(51,90){\line(1,0){88}}
\put(141,90){\line(1,0){9}}
\put(170,90){\line(1,0){10}}
\put(10,100){\line(1,0){10}}
\put(40,100){\line(1,0){110}}
\put(170,100){\line(1,0){10}}
\put(20,85){\framebox(20,20){$W$}}
\put(20,60){\framebox(20,20){$W$}}
\put(20,20){\framebox(20,20){$W$}}
\put(150,85){\framebox(20,20){$W^\dag$}}
\put(150,60){\framebox(20,20){$W^\dag$}}
\put(150,20){\framebox(20,20){$W^\dag$}}
\put(50,5){\circle{4}}
\put(50,90){\circle{2}}
\put(50,100){\circle*{2}}
\put(50,100){\line(0,-1){9}}
\put(50,89){\line(0,-1){86}}
\put(60,5){\circle{4}}
\put(60,65){\circle{2}}
\put(60,75){\circle*{2}}
\put(60,75){\line(0,-1){9}}
\put(60,64){\line(0,-1){61}}
\put(80,5){\circle{4}}
\put(80,25){\circle{2}}
\put(80,35){\circle*{2}}
\put(80,35){\line(0,-1){9}}
\put(80,24){\line(0,-1){21}}
\put(110,5){\circle{4}}
\put(110,25){\circle{2}}
\put(110,35){\circle*{2}}
\put(110,35){\line(0,-1){9}}
\put(110,24){\line(0,-1){21}}
\put(130,5){\circle{4}}
\put(130,65){\circle{2}}
\put(130,75){\circle*{2}}
\put(130,75){\line(0,-1){9}}
\put(130,64){\line(0,-1){61}}
\put(140,5){\circle{4}}
\put(140,90){\circle{2}}
\put(140,100){\circle*{2}}
\put(140,100){\line(0,-1){9}}
\put(140,89){\line(0,-1){86}}
\put(95,15){\circle{2}}
\put(95,14){\line(0,-1){4}}
\put(88,0){\framebox(14,10){$e^{-iZt}$}}
\put(30,45){\circle*{1}}
\put(30,50){\circle*{1}}
\put(30,55){\circle*{1}}
\put(68,55){\circle*{1}}
\put(70,50){\circle*{1}}
\put(72,45){\circle*{1}}
\put(160,45){\circle*{1}}
\put(160,50){\circle*{1}}
\put(160,55){\circle*{1}}
\put(122,55){\circle*{1}}
\put(120,50){\circle*{1}}
\put(118,45){\circle*{1}}
\put(0,0){\makebox(10,10){$|0\rangle$}}
\put(0,10){\makebox(10,10){$r$}}
\put(0,20){\makebox(10,10){$b_{2n}$}}
\put(0,30){\makebox(10,10){$a_{2n}$}}
\put(0,60){\makebox(10,10){$b_2$}}
\put(0,70){\makebox(10,10){$a_2$}}
\put(0,85){\makebox(10,10){$b_1$}}
\put(0,95){\makebox(10,10){$a_1$}}
\end{picture}
\caption{A circuit for simulating $e^{-iTt}$.}
\label{fig:simT}
\end{figure}
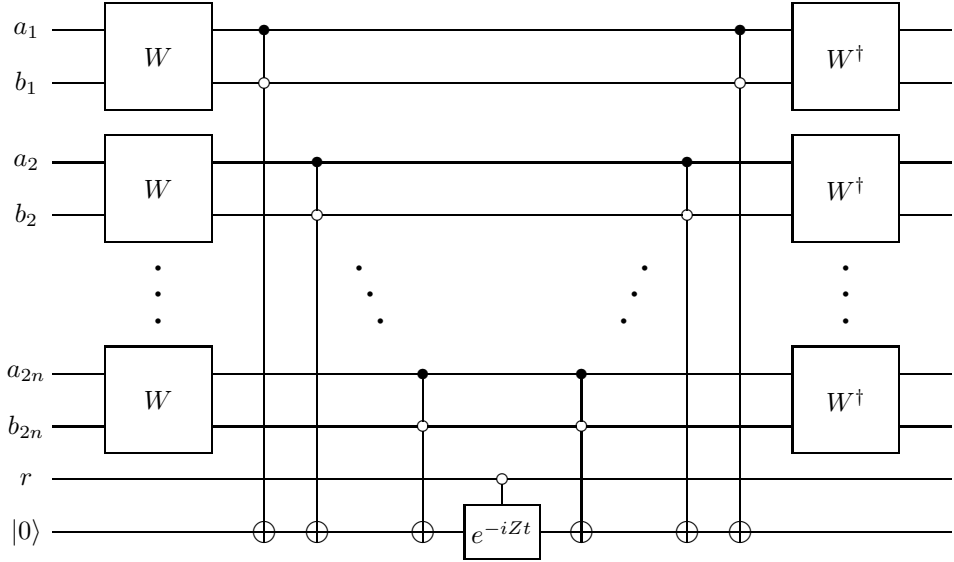

The operator $T$ may be written as
\be
  T = \left( \bigotimes_{l=1}^{2n} S^{(l,2n+l)} \right) \otimes |0\>\<0|
\label{eq:swap}
\ee
where the superscript indicates which two qubits $S$ acts on, and the
projector onto $|0\>$ acts on the third register.  Here $S$ is a Hermitian
operator on two qubits satisfying $S |z_1 z_2\> = |z_2 z_1\>$.  Since the
eigenvalues of $S$ are $\pm1$, the eigenvalues of $T$ are $0,\pm1$, and
they are easy to compute.  Thus $e^{-i T t}$ can be simulated with the
circuit shown in Figure~\ref{fig:simT}.  In this figure, $W$ denotes a
two-qubit unitary operator that diagonalizes $S$.  The unique eigenvector
of $S$ with eigenvalue $-1$ is ${1 \over \sqrt2}(|01\>-|10\>)$, so we take
\bea
  W |00\> &=& |00\> \\
  W {\textstyle{1 \over \sqrt2}}(|01\> + |10\>) &=& |01\> \\
  W {\textstyle{1 \over \sqrt2}}(|01\> - |10\>) &=& |10\> \\
  W |11\> &=& |11\>
\,.
\eea
Applying $W^{\otimes 2n}$ diagonalizes $T$, and the Toffoli gates compute
the argument of the eigenvalue in an ancilla register initially prepared
in the state $|0\>$.  Note that a filled circle denotes ``control on
$|1\>$'' and an open circle denotes ``control on $|0\>$.''  After
computing the eigenvalue, we apply the appropriate phase shift if $r=0$ by
evolving for a time $t$ according to the Pauli $Z$ operator satisfying
$Z|z\>=(-1)^z |z\>$.  The controlled phase shift can be performed since it
is a local term.  Finally, we uncompute the eigenvalue and return to the
original basis.

Our Hamiltonian for the quantum walk on $G$ is
\be
  H = \sum_c V_c^\dag T V_c
\label{eq:simham}
\ee
where the sum runs over all $k$ colors.  Given the simulation of $T$
described above, this can be simulated using unitary conjugation and
linear combination.  Note that $V_c^\dag = V_c$, so we do not need to
separately explain how to perform $V_c^\dag$.

To conclude, we show that (\ref{eq:simham}) acts as it should to satisfy
(\ref{eq:ham}).  We have
\bea
  H |a, 0, 0\>
  &=& \sum_c V_c T |a, v_c(a), f_c(a)\> \\
  &=& \sum_c \delta_{0,f_c(a)} V_c |v_c(a), a, 0\> \\
  &=& \sum_{c:\, v_c(a) \in G} |v_c(a), a \oplus v_c(v_c(a)), f_c(v_c(a))\>
\,.
\eea
Noting that $v_c(a) \in G \Rightarrow f_c(v_c(a))=0$ and $v_c(v_c(a))=a$,
this shows that
\be
  H |a, 0, 0\> = \sum_{c:\, v_c(a) \in G} |v_c(a),0,0\>
\ee
as desired.

Although the continuous time quantum walk is most naturally viewed as
Schr\"odinger evolution according to a time-independent Hamiltonian, we
have shown how to simulate this evolution in the conventional circuit
model of quantum computation.  Note that since the names of the vertices
may be arbitrary, the graph Hamiltonian may not be local, but nevertheless
it can be simulated in the circuit model.

%%%%%%%%%%%%%%%%%%%%%%%%%%%%%%%%%%%%%%%%%%%%%%%%%%%%%%%%%%%%%%%%%%%%%%%%%%%%%%%
\subsection{Propagation on a line}\label{subsec:line}

We now consider the quantum walk on the specific family of graphs $G_n'$
of Figure \ref{fig:graphprime}.  In this section, we use standard physics
techniques to give compelling evidence that the quantum walk propagates
from the $\ent$ to the $\exit$ in linear time.  We explain how the walk on
$G_n'$ can be viewed as a walk on a finite line with a defect at the
center, and we show through a series of examples that the defect and the
boundaries do not significantly affect the walk.  In section
\ref{subsec:hitting}, we prove that the walk reaches the $\exit$ in
polynomial time.

\begin{figure}
\setlength{\unitlength}{1.85pt}
\begin{tabular}{r@{\hspace{36pt}}c}
\raisebox{16pt}{(a)} &
\begin{picture}(220,20)
\put(10,10){\circle*{2}}
\put(30,10){\circle*{2}}
\put(50,10){\circle*{2}}
\put(60,10){\circle*{.5}}
\put(65,10){\circle*{.5}}
\put(70,10){\circle*{.5}}
\put(80,10){\circle*{2}}
\put(100,10){\circle*{2}}
\put(120,10){\circle*{2}}
\put(140,10){\circle*{2}}
\put(150,10){\circle*{.5}}
\put(155,10){\circle*{.5}}
\put(160,10){\circle*{.5}}
\put(170,10){\circle*{2}}
\put(190,10){\circle*{2}}
\put(210,10){\circle*{2}}
\put(10,10){\line(+1,0){45}}
\put(75,10){\line(+1,0){70}}
\put(165,10){\line(+1,0){45}}
\put(15,10){\makebox(10,10){$1$}}
\put(35,10){\makebox(10,10){$1$}}
\put(85,10){\makebox(10,10){$1$}}
\put(105,10){\makebox(10,10){$\sqrt 2$}}
\put(125,10){\makebox(10,10){$1$}}
\put(175,10){\makebox(10,10){$1$}}
\put(195,10){\makebox(10,10){$1$}}
\put(-13,5){\makebox(20,10){\scriptsize $\ent$}}
\put(212,5){\makebox(10,10){\scriptsize $\exit$}}
\put(5,0){\makebox(10,10){\scriptsize $\col 0$}}
\put(25,0){\makebox(10,10){\scriptsize $\col 1$}}
\put(45,0){\makebox(10,10){\scriptsize $\col 2$}}
\put(75,0){\makebox(10,10){\scriptsize $\col n\!-\!1$}}
\put(95,0){\makebox(10,10){\scriptsize $\col n$}}
\put(115,0){\makebox(10,10){\scriptsize $\col n\!+\!1$}}
\put(135,0){\makebox(10,10){\scriptsize $\col n\!+\!2$}}
\put(165,0){\makebox(10,10){\scriptsize $\col 2n\!-\!1$}}
\put(185,0){\makebox(10,10){\scriptsize $\col 2n$}}
\put(205,0){\makebox(10,10){\scriptsize $\col 2n\!+\!1$}}
\end{picture}
\vspace{10pt}
\\ \raisebox{16pt}{(b)} &
\begin{picture}(160,20)
\put(10,10){\circle*{2}}
\put(30,10){\circle*{2}}
\put(50,10){\circle*{2}}
\put(70,10){\circle*{2}}
\put(90,10){\circle*{2}}
\put(110,10){\circle*{2}}
\put(130,10){\circle*{2}}
\put(150,10){\circle*{2}}
\put(10,10){\vector(+1,0){150}}
\put(10,10){\vector(-1,0){10}}
\put(15,10){\makebox(10,10){$1$}}
\put(35,10){\makebox(10,10){$1$}}
\put(55,10){\makebox(10,10){$1$}}
\put(75,10){\makebox(10,10){$1$}}
\put(95,10){\makebox(10,10){$1$}}
\put(115,10){\makebox(10,10){$1$}}
\put(135,10){\makebox(10,10){$1$}}
\end{picture}
\vspace{10pt}
\\ \raisebox{16pt}{(c)} &
\begin{picture}(160,20)
\put(10,10){\circle*{2}}
\put(30,10){\circle*{2}}
\put(50,10){\circle*{2}}
\put(70,10){\circle*{2}}
\put(90,10){\circle*{2}}
\put(110,10){\circle*{2}}
\put(130,10){\circle*{2}}
\put(150,10){\circle*{2}}
\put(10,10){\vector(+1,0){150}}
\put(10,10){\vector(-1,0){10}}
\put(65,0){\makebox(10,10){\scriptsize $j=d$}}
\put(85,0){\makebox(10,10){\scriptsize $j=d\!+\!1$}}
\put(15,10){\makebox(10,10){$1$}}
\put(35,10){\makebox(10,10){$1$}}
\put(55,10){\makebox(10,10){$1$}}
\put(75,10){\makebox(10,10){$\alpha$}}
\put(95,10){\makebox(10,10){$1$}}
\put(115,10){\makebox(10,10){$1$}}
\put(135,10){\makebox(10,10){$1$}}
\end{picture}
\vspace{10pt}
\\ \raisebox{16pt}{(d)} &
\begin{picture}(170,20)
\put(10,10){\circle*{2}}
\put(30,10){\circle*{2}}
\put(50,10){\circle*{2}}
\put(70,10){\circle*{2}}
\put(80,10){\circle*{.5}}
\put(85,10){\circle*{.5}}
\put(90,10){\circle*{.5}}
\put(100,10){\circle*{2}}
\put(120,10){\circle*{2}}
\put(140,10){\circle*{2}}
\put(160,10){\circle*{2}}
\put(10,10){\line(+1,0){65}}
\put(95,10){\line(+1,0){65}}
\put(5,0){\makebox(10,10){\scriptsize $j=1$}}
\put(155,0){\makebox(10,10){\scriptsize $j=L$}}
\put(15,10){\makebox(10,10){$1$}}
\put(35,10){\makebox(10,10){$1$}}
\put(55,10){\makebox(10,10){$1$}}
\put(105,10){\makebox(10,10){$1$}}
\put(125,10){\makebox(10,10){$1$}}
\put(145,10){\makebox(10,10){$1$}}
\end{picture}
\end{tabular}
\caption{Quantum walks on lines.
  (a) Reduction of the quantum walk on $G_n'$ to a quantum walk on a line.
  (b) Quantum walk on an infinite, translationally invariant line.
  (c) Quantum walk on an infinite line with a defect.
  (d) Quantum walk on a finite line without a defect.}
\label{fig:line}
\end{figure}
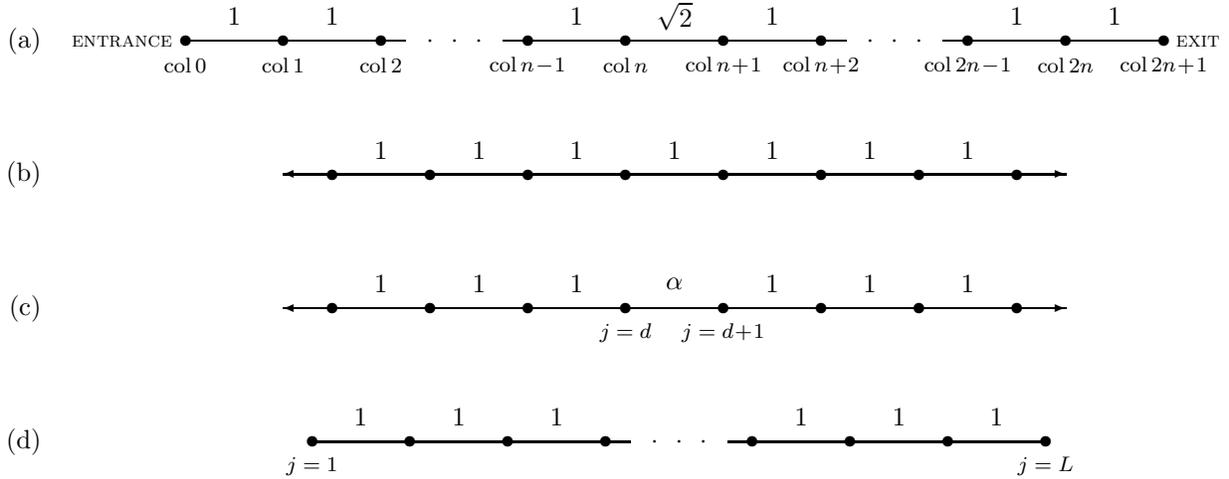

The analysis of the walk on $G_n'$ is particularly simple because it can
be reduced to a walk on a line with $2n+2$ vertices, one for each column
of the original graph.  Consider the $(2n+2)$-dimensional subspace spanned
by the states
\be
  |\col j\> = {1 \over \sqrt N_j} \sum_{a \in \mathrm{column~} j} |a\>
\,,
\ee
where
\be
  N_j = \cond{2^j        &   0 \le j \le n \\
              2^{2n+1-j} & n+1 \le j \le 2n+1 \,.}
\ee
We refer to this subspace as the {\em column subspace}. Because every vertex
in column $j$ is connected to the same number of vertices in column $j+1$
and every vertex in column $j+1$ is connected to the same number of vertices
in column $j$, applying $H$ to any state in
the column subspace results in another state in this subspace.
Despite the random connections in $G^{\prime}_n$, the column subspace is 
invariant under $H$.
In particular, a quantum walk starting in the state corresponding to the
$\ent$ always remains in the column subspace.  Thus, to understand the
quantum walk starting from the $\ent$, we only need to understand how the
Hamiltonian acts on the column subspace.  In this subspace, the non-zero
matrix elements of $H$ are
\be
  \<\col j|H|\col(j+1)\> = \cond{
   \sqrt 2 \gamma & 0 \le j \le n-1 \,,~~ n+1 \le j \le 2n \\
   2 \gamma       & j=n}
\label{eq:lineham}
\ee
(and those deduced by Hermiticity of $H$).  For simplicity, we set
$\gamma=1/\sqrt2$.  The quantum walk in the column subspace is shown
pictorially in Figure \ref{fig:line}(a).

We claim that if the quantum state at $t=0$ is $|\col 0\>$, then at a time
of order $n/2$, there is an appreciable probability of being at
$|\col(2n+1)\>$.  To see this, first consider propagation on an infinite,
translationally invariant line as shown in Figure \ref{fig:line}(b).  The
nonzero matrix elements of the Hamiltonian are
\be
  \<j|H|j\pm 1\> = 1 \,,\quad -\infty < j < \infty
\,.
\ee
The eigenstates of this Hamiltonian
are the momentum eigenstates $|p\>$ with components
\be
  \<j|p\> = {1 \over \sqrt{2\pi}} e^{i p j}
  \,, \quad -\pi \le p \le \pi
\ee
having energies
\be
  E_p = 2 \cos p 
\,.
\ee
The propagator, or Green's function, to go from $j$ to $k$ in time $t$ is
\bea
\label{eq:green}
  G(j,k,t) &=& \<k|e^{-i H t}|j\> \\
           &=& {1 \over 2 \pi} \int_{-\pi}^\pi \! \d p \, 
	       e^{i p (k-j) - 2 i t \cos p} \\
	   &=& (-i)^{k-j} J_{k-j}(2t) \,,\quad -\infty < j,k < \infty
\eea
where $J_\nu(\cdot)$ is a Bessel function of order $\nu$.  By the
well-known properties of the Bessel function, this shows that a state
initially localized in a single column evolves as a left moving and a
right moving wave packet, each propagating with speed $2$.  To see this,
note that the Bessel function has the following asymptotic expansions for
$\nu \gg 1$:
\bea
  J_\nu(\nu \sech \xi) 
    &\sim& {e^{-\nu(\xi-\tanh \xi)} \over \sqrt{2 \pi \nu \tanh \xi}} 
    \label{eq:besselahead} \\
  J_\nu(\nu+\xi\nu^{1/3}) 
    &=& (2/\nu)^{1/3} \Ai(-2^{1/3}\xi) + O(\nu^{-1})
    \label{eq:besseledge} \\
  J_\nu(\nu \sec \xi)
       &=& \sqrt{2 \over \pi \nu \tan\xi} \left\{
           \cos[\textstyle{\pi \over 4}-\nu(\xi-\tan \xi)] + O(\nu^{-1})
	   \right\}
           \,, \quad 0 < \xi < {\pi \over 2} \label{eq:besselbehind}
\eea
where $\Ai(\cdot)$ is an Airy function \cite{AS72}.  These three relations
show that for $|k-j| \gg 1$, $G(j,k,t)$ is exponentially small in $|k-j|$
for $t <0.99\cdot |k-j|/2$, of order $|k-j|^{-1/3}$ for $t$ near $|k-j|/2$,
and of order $|k-j|^{-1/2}$ for $t >1.01\cdot |k-j|/2$.

\begin{figure}
\includegraphics[width=3.3in]{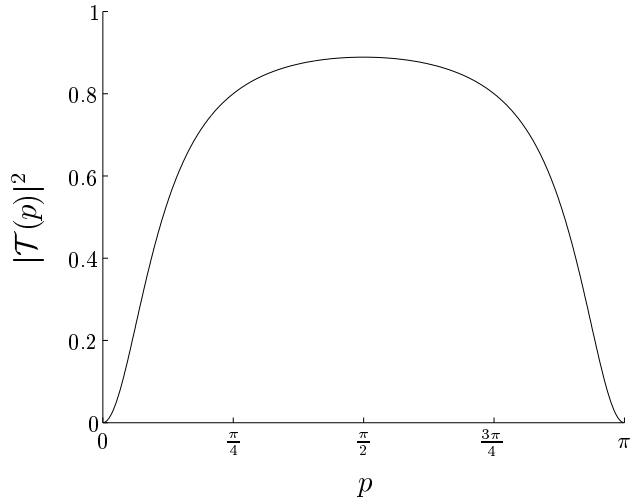}
\caption{The transmission probability $|{\cal T}(p)|^2$ as a function of
momentum for the infinite line with a defect (with $\alpha=\sqrt 2$).}
\label{fig:transmit}
\end{figure}

To understand the effect of a defect, we now consider an infinite line
with a defect between sites $j=d$ and $j=d+1$, as shown in
\ref{fig:line}(c).  For generality, we consider the case where the matrix
element across the defect is $\alpha$.  We use standard scattering theory
to calculate the transmission coefficient for an incident plane wave of
momentum $p>0$.  Consider a state $|\psi\>$ consisting of an incident
plane wave and a reflected plane wave on the left side of the defect and a
transmitted plane wave on the right side:
\be
  \<j|\psi\> = \cond{
  {1 \over \sqrt{2\pi}} e^{i p j}
  +{{\cal R} \over \sqrt{2\pi}} e^{-i p j} & j \le d \\
   {{\cal T} \over \sqrt{2\pi}} e^{i p j}  & j \ge d+1 \,.}
\ee
Here $\cal R$ is the reflection coefficient and $\cal T$ is the
transmission coefficient.  Requiring this to be an eigenstate of the
Hamiltonian, we find
\be
  {\cal T}(p) 
    = {2i \alpha \sin p \over (\alpha^2-1) \cos p + i (\alpha^2+1) \sin p}
\ee
which gives
\be
  |{\cal T}(p)|^2 
    = {8 \sin^2 p \over 1 + 8 \sin^2 p}
\ee
for $\alpha=\sqrt 2$, as shown in Figure \ref{fig:transmit}. A
narrow wave packet with momentum $p$ propagates through the defect
with probability $|{\cal T}(p)|^2$.  The wave packet that results from a
state initially localized at a particular site is spread out over a range
of momenta, but since the transmission probability is appreciable over
most of the range, it is clear that there will be substantial transmission
through the defect.

Finally, to show that boundaries do not impede the propagation, we now
consider the finite line (without a defect) shown in Figure
\ref{fig:line}(d).  This problem can be treated by standard techniques of
multiple reflection.  The exact Green's function $\tilde G(j,k,t)$ for
this finite line in terms of the Green's function $G(j,k,t)$ for the
infinite line is\footnote{Note that (\ref{eq:green2}) is the exact formula
for propagation in the graph $G_n$ shown in Figure \ref{fig:graph} and
studied in \cite{CFG02} if we use (\ref{eq:graphham}) instead of
(\ref{eq:graphham1}) as the definition of a quantum walk.}
\be
\label{eq:green2}
  \tilde G(j,k,t) = 
  \sum_{l=-\infty}^\infty \left[G(j,k+2l(L+1),t) -
  G(j,-k+2l(L+1),t)\right]
\,, \quad
  1 \le j,k \le L
\,.
\ee
This can be interpreted as a sum of paths making reflections off the
boundaries.  To verify this formula, the reader can check that it
satisfies the Schr\"odinger equation, the boundary conditions, and the
initial condition.  The Schr\"odinger equation is satisfied because each
term individually satisfies it for the infinite line.  The boundary
conditions $\tilde G(j,0,t)=\tilde G(j,L+1,t)=0$ can be verified by
substitution.  The initial condition $\tilde G(j,k,0)=\delta_{jk}$ holds
because $G(j,k,0)=\delta_{jk}$, and the only contribution at $t=0$ comes
from the first term of (\ref{eq:green2}) with $l=0$.

We will now see, using the particular form of the propagator in terms of a
Bessel function, that propagation from $j=1$ to $j=L$ takes time $L/2$ for
$L \gg 1$.  Since $J_{k-j}(2t)$ is small for $|k-j| \gg 2t$, there are
only four terms that contribute significantly for $t \gtrsim L/2$.  They
result from taking $l=0,-1$ in the first term of (\ref{eq:green2}) and
$l=0,1$ in the second term.  The resulting expression is
\be
  \tilde G(1,L,t) \approx G(1,L,t)+G(1,-L-2,t)-G(1,-L,t)-G(1,L+2,t)
  \,, \quad t \gtrsim L/2
\,,
\ee
the magnitude of which is not small.

We have seen that propagation on a line occurs as a wave packet with speed
$2$, that the wave packet is substantially transmitted through a defect,
and that reflections off boundaries do not impede propagation.  Taken
together, these facts constitute compelling evidence that the quantum walk
traverses the graph $G_n'$ in linear time.  The exact propagator for the
line shown in Figure \ref{fig:line}(a) can be calculated using a more
sophisticated version of these techniques \cite{GoldComm}.  This exact
propagator, evaluated for $t$ near $n$, is of order $n^{-1/3}$.  We spare
the reader the details of this calculation and give a simpler proof of a
bound that is not tight---but is nevertheless polynomial---in the
following section.

%%%%%%%%%%%%%%%%%%%%%%%%%%%%%%%%%%%%%%%%%%%%%%%%%%%%%%%%%%%%%%%%%%%%%%%%%%%%%%%
\subsection{Upper bound on the hitting time}\label{subsec:hitting}

Although the preceding section demonstrated beyond any reasonable doubt
that the quantum walk traverses the graph $G_n'$ in linear time, we now
provide a simple proof that the hitting time is upper bounded by a
polynomial.

For the purpose of this proof, it will be more convenient to consider the
graph $G_{n-1}'$, which reduces to a line with $2n$ vertices.  We label
the vertices from $1$ to $2n$, and the defect is on the edge between
vertices $n$ and $n+1$.  With this labeling and $\gamma=1/\sqrt2$, the
Hamiltonian (\ref{eq:lineham}) is
\be
  \<\col j|H|\col(j+1)\> = \cond{
   1      & 1 \le j \le n-1 \,,~~ n+1 \le j \le 2n-1 \\
   \sqrt2 & j=n \,,}
\ee
with Hermiticity of $H$ giving the other nonzero matrix elements.

Define a reflection operator
\be
  R|\col j\> = |\col(2n+1-j)\>
\,.
\ee
Note that $R^2=1$, so $R$ has eigenvalues $\pm 1$.  $R$ commutes with $H$
on the column subspace, so we can find simultaneous eigenstates of $R$ and
$H$.  These are of the form
\be
  \<\col j|E\> = \cond{
    \sin p j          & 1 \le j \le n \\
    \pm\sin(p(2n+1-j)) & n+1 \le j \le 2n \,, }
\label{eq:eigenstates}
\ee
which explicitly vanish at $j=0$ and $j=2n+1$.  The eigenvalue
corresponding to the eigenstate $|E\>$ is $E=2 \cos p$, and the
quantization condition (to be discussed later) comes from matching at
vertices $n$ and $n+1$.  The $\ent$ vertex corresponds to $|\col 1\>$ and
the $\exit$ vertex to $|\col 2n\>$.

\begin{lemma}\label{lemma:hitting}
Consider the quantum walk in $G_{n-1}'$ starting at the $\ent$.  Let the
walk run for a time $t$ chosen uniformly in $[0,\tau]$ and then measure in
the computational basis.  If $\tau \ge {4n \over \epsilon \, \Delta E}$
for any constant $\epsilon>0$, where $\Delta E$ is the magnitude of the
smallest gap between any pair of eigenvalues of the Hamiltonian, then the
probability of finding the $\exit$ is greater than ${1 \over
2n}(1-\epsilon)$.
\end{lemma}

\begin{proof}
The probability of finding the $\exit$ after the randomly chosen time $t
\in [0,\tau]$ is
\bea
  && {1 \over \tau} \int_0^\tau \! \d t \,
     |\<\col 2n|e^{-i H t}|\col 1\>|^2 \nonumber\\
  &&\qquad = \, {1 \over \tau} \sum_{E,E'} \int_0^\tau \! \d t \, 
                e^{-i(E-E')t} \<\col 2n|E\>\<E|\col 1\> 
                              \<\col 1|E'\>\<E'|\col 2n\> \\
  &&\qquad = \, \sum_E |\<E|\col 1\>|^2 |\<E|\col 2n\>|^2 \nonumber\\
  &&\qquad ~ \,\, + \sum_{E \ne E'} {1 - e^{-i(E-E')\tau} \over i(E-E')\tau}
	      \<\col 2n|E\>\<E|\col 1\> \<\col 1|E'\>\<E'|\col 2n\>
\,.
\eea
Because of (\ref{eq:eigenstates}), we have $\<E|\col 1\>=\pm\<E|\col
2n\>$.  Thus the first term is
\be
  \sum_E |\<E|\col 1\>|^4 \ge {1 \over 2n}
\ee
as is easily established using the Cauchy-Schwartz inequality.  The second
term can be bounded as follows:
\bea
  && \left |\sum_{E \ne E'} {1 - e^{-i(E-E')\tau} \over i(E-E')\tau}
           \<\col 2n|E\>\<E|\col 1\> 
           \<\col 1|E'\>\<E'|\col 2n\> \right| \nonumber\\
  && \qquad \le {2 \over \tau \Delta E} 
     \sum_{E,E'} |\<E|\col 1\>|^2 |\<E'|\col 2n\>|^2
    ={2 \over \tau \Delta E}
\,.
\eea
Thus we have
\be
  {1 \over \tau} \int_0^\tau \! \d t \, |\<\col 2n|e^{-i H t}|\col 1\>|^2
  \ge {1 \over 2n} - {2 \over \tau \Delta E}
  \ge {1 \over 2n} (1-\epsilon)
\ee
where the last inequality follows since $\tau \ge {4 n \over \epsilon \,
\Delta E}$ by assumption.
\end{proof}

Now we need to prove that the minimum gap $\Delta E$ is only polynomially
small.

\begin{lemma}\label{lemma:gap}
The smallest gap between any pair of eigenvalues of the Hamiltonian
satisfies
\be
  \Delta E > {2 \pi^2 \over (1+\sqrt2)n^3} + O(1/n^4)
\,.
\ee
\end{lemma}

\begin{proof}
To evaluate the spacings between eigenvalues, we need to use the
quantization condition.  We have
\be
  \<\col n|H|E\> = 2 \cos p \, \<\col n|E\>
\ee
so that
\be
  \sqrt2 \<\col(n+1)|E\> + \<\col(n-1)|E\> = 2 \cos p \, \<\col n|E\>
\ee
and using (\ref{eq:eigenstates}), we have
\be
  \pm \sqrt2 \sin np + \sin((n-1)p) = 2 \cos p \, \sin np
\ee
which simplifies to
\be
  {\sin((n+1)p) \over \sin np} = \pm \sqrt 2
\,.
\label{eq:constraint}
\ee
In Figure \ref{fig:constraint} we plot the left hand side of
(\ref{eq:constraint}) for $n=5$.  The intersections with $-\sqrt2$ occur
to the left of the zeros of $\sin np$, which occur at $\pi l/n$ for
$l=1,2,\ldots,n-1$.  For the values of $p$ that intersect $-\sqrt 2$, we
can write $p=(\pi l/n)-\delta$.  Equation (\ref{eq:constraint}) with
$-\sqrt2$ on the right hand side is now
\be
  -\sqrt2 \sin n\delta = \sin\left(n\delta - {l\pi \over n} + \delta\right)
\,.
\label{eq:minus}
\ee
Write $\delta = (c/n) + (d/n^2) + O(1/n^3)$.  Taking $n\to\infty$ in
(\ref{eq:minus}) gives $-\sqrt2 \sin c = \sin c$, which implies that
$c=0$.  We then get
\be
  -\sqrt2 \sin\left({d \over n}+O(1/n^2)\right) 
        = \sin\left({d \over n}-{l \pi \over n}+O(1/n^2)\right)
\ee
which gives, as $n\to\infty$,
\be
  d = {l \pi \over 1+\sqrt 2}
\,.
\ee
Thus we have that the roots of (\ref{eq:constraint}) with $-\sqrt2$ on the
right hand side are of the form
\be
  p = {l \pi \over n} - {l \pi \over (1+\sqrt 2)n^2} + O(1/n^3)
\,.
\ee

\begin{figure}
\includegraphics[width=3.3in]{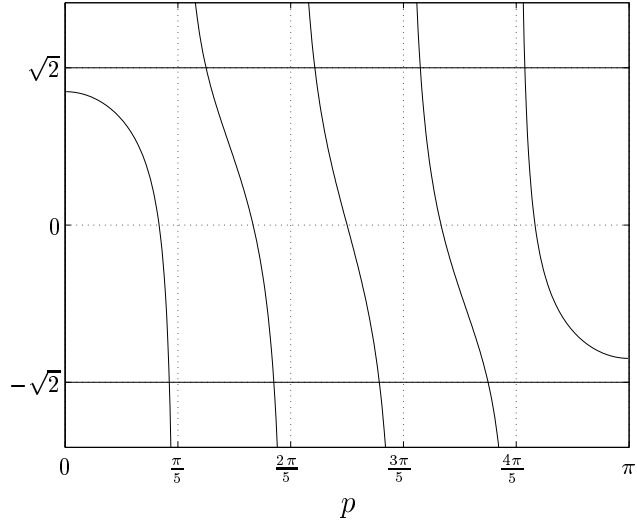}
\caption{Left hand side of (\ref{eq:constraint}) for $n=5$.}
\label{fig:constraint}
\end{figure}

Let $p'$ and $p''$ be the two roots of (\ref{eq:constraint}) closest to
the root $p$ just found, with $p' < p < p''$.  From the figure we see that
$p'$ and $p''$ both are roots of (\ref{eq:constraint}) with $+\sqrt2$.
(Note that the smallest $p$, corresponding to $l=1$, does not have a
$p'$.)   We see that $p''$ lies to the right of the zero of $\sin np$ at
$p=l\pi/n$.  We also see that $p'$ lies to the left of the zero of
$\sin((n+1)p)$ at $l \pi/(n+1)$.  Therefore we have
\bea
  p'  &<& {l \pi \over n} - {l \pi \over n^2} + O(1/n^3) \\
  p'' &>& {l \pi \over n}
\,,
\eea
from which we conclude that
\bea
  p-p'  &>& {l \pi \sqrt 2 \over (1+\sqrt2)n^2} + O(1/n^3) 
            \,, \quad l=2,3,\ldots,n-1 \\
  p''-p &>& {l \pi \over (1+\sqrt2)n^2} + O(1/n^3)
            \,, \quad l=1,2,\ldots,n-1
\,.
\eea
Thus the smallest spacing is at least $\pi/[(1+\sqrt2)n^2] + O(1/n^3)$.

Now for a given $p$, the corresponding eigenvalue is $2\cos p$.  For small
$\Delta p$, the spacing $\Delta E$ is related to the spacing $\Delta p$ by
\be
  \Delta E = 2 |\Delta p \, \sin p| + O\left((\Delta p)^2\right)
\,.
\ee
The factor $\sin p = \sin(l\pi/n + O (1/n^2))$ is smallest when $l=1$, so
we have
\be
  \Delta E > {2 \pi^2 \over (1+\sqrt 2) n^3} + O(1/n^4) > {8 \over n^3}
  \textrm{~ for $n$ sufficiently large.}
\ee

The alert reader will note from the figure with $n=5$ that there are only
$8$ roots, whereas the dimension of the reduced space is $10$, so there
are actually $10$ eigenvalues.  In general, there are $n-2$ roots of
(\ref{eq:constraint}) with $p$ real.  If we let $p=ik$ with $k$ real, we
can have eigenstates of the form (\ref{eq:eigenstates}) with $\sin pj$
replaced by $\sinh kj$ and $\pm \sin(p(2n+1-j))$ replaced by
$\pm\sinh(k(2n+1-j))$.  The corresponding eigenvalue is $2 \cosh k$ and
the condition (\ref{eq:constraint}) becomes
\be
  {\sinh((n+1)k) \over \sinh nk} = \pm \sqrt2 .
\label{cond}
\ee
As $n\to\infty$, the root of this equation is at $e^k=\sqrt{2}$, which
corresponds to an eigenvalue $\sqrt 2+{1 \over \sqrt 2}$. To obtain the
last eigenvalue, let $p=\pi + ik$.  The eigenvalue is then $-2\cosh k$.
The quantization condition is now the same as (\ref{cond}), and as
$n\to\infty$, the eigenvalue is $-(\sqrt 2+{1 \over \sqrt 2})$.  So we
have found two eigenvalues at $\pm(\sqrt 2+{1 \over \sqrt 2})$ with
corrections that vanish exponentially as $n\to\infty$.  Since the other
$n-2$ eigenvalues are all in the range $[-2,2]$, our conclusion about the
minimum spacing is unchanged.
\end{proof}

Using Lemma \ref{lemma:hitting} and Lemma \ref{lemma:gap}, we find
\begin{theorem}
For $n$ sufficiently large, running the quantum walk for a time chosen
uniformly in $[0,{n^4 \over 2\epsilon}]$ and then measuring in the
computational basis yields a probability of finding the $\exit$ that is
greater than ${1 \over 2n} (1-\epsilon)$.
\label{thm:hitting}
\end{theorem}

To summarize, we have presented an efficient algorithm for traversing any
graph $G_n'$ using a quantum computer.  The computer is prepared in the
state corresponding to the $\ent$, and the quantum walk is simulated using
the construction described in Section \ref{subsec:implement}.  After
running the walk for a certain time $t$, the state of the computer is
measured in the computational basis.  The oracle can then be used to check
whether the resulting vertex name corresponds to a vertex of degree 2
other than $\ent$, in which case it must be $\exit$.  Theorem
\ref{thm:hitting} shows that by choosing an appropriate $t=\poly(n)$, the
probability of finding the name of the $\exit$ can be $O(1/n)$.  By
repeating this process $\poly(n)$ times, the success probability can be
made arbitrarily close to 1.  Combining this with the efficient
implementation of the quantum walk described in Section
\ref{subsec:implement}, we see that the quantum walk algorithm finds the
name of the $\exit$ with high probability using $\poly(n)$ calls to the
oracle.  

%%%%%%%%%%%%%%%%%%%%%%%%%%%%%%%%%%%%%%%%%%%%%%%%%%%%%%%%%%%%%%%%%%%%%%%%%%%%%%%
\section{Classical lower bound}\label{sec:lowerbound}

We now prove that no classical algorithm can find the $\exit$ with high
probability in subexponential time.  We do this by considering a series of
games and proving relations between them.  The first game is equivalent to
our problem, and each new game will be essentially as easy to win.
Finally, we will show that the easiest game cannot be won in
subexponential time.

Until now, we have not defined the specific coloring of the edges of the
graph.  We did not need to consider a particular coloring in Section
\ref{sec:algorithm} because the quantum algorithm works given {\em any}
consistent coloring.  However, to prove the classical lower bound, we need
to specify a coloring that does not supply information about the graph.
The coloring is chosen at random as follows.  At each vertex in an even
numbered column, randomly color the incident edges $A,B,C$.  At each
vertex in an odd numbered column, randomly append $1,2,3$ to the colors of
the incident edges.  The edges are then colored with one of nine colors,
$A1,A2,A3,B1,\ldots,C3$.  Because of the structure of $G_n'$, any such
coloring is consistent, but as we will see below, it does not provide any
useful information to a classical algorithm.

Our problem is equivalent to the following game:
\begin{game}\label{game:original}
  The oracle contains a random set of names for the vertices of the
  randomly chosen graph $G_n'$ such that each vertex has a distinct
  $2n$-bit string as its name and the $\ent$ vertex has the name $0$.  In
  addition, the edges of the graph are randomly colored as described
  above.
  At each step, the algorithm sends a $2n$-bit string to the oracle, and
  if there exists a vertex with that name, the oracle returns the names of
  the neighbors of that vertex and the colors of the edges that join
  them.
  The algorithm wins if it ever sends the oracle the name of the $\exit$
  vertex.
\end{game}

\noindent
Note that there are three sources of randomness in this oracle: in the
choice of a graph $G_n'$, in the random naming of its vertices, and in the
random coloring of its edges.  We first consider a fixed graph $G$ and
coloring $C$ and only draw implications from the random names.  Throughout
this section, $G$ always refers to one of the graphs $G_n'$.  For a game
$X$ with a graph $G$ and a coloring $C$, the success probability of the
algorithm $A$ is defined as
\be
  \mathbb{P}_X^{G,C}(A) =
  \prob{\mathrm{names}}
       {\mbox{\rm algorithm $A$ wins game $X$ on graph $G$ with coloring $C$}}
\ee
where $\displaystyle\prob{\mathrm{names}}{\cdot}$ means the probability is taken over
the random naming of vertices.

In Game \ref{game:original}, the algorithm could traverse a disconnected
subgraph of $G_n'$.  But because there are exponentially many more strings
of $2n$ bits than vertices in the graph, it is highly unlikely that any
algorithm will ever guess the name of a vertex that it was not sent by the
oracle.  Thus, Game \ref{game:original} is essentially equivalent to the
following game:

\begin{game}\label{game:tree}
  The oracle contains a graph, set of vertex names, and edge coloring as
  described in Game \ref{game:original}.
  At each step, the algorithm sends the oracle the name of the $\ent$
  vertex or the name of a vertex it has previously been sent by the
  oracle.  The oracle then returns the names of the neighbors of that
  vertex and the colors of the edges that join them.
  The algorithm wins it ever sends the oracle the name of the $\exit$
  vertex.
\end{game}

\noindent
The next lemma shows that, if the algorithms run for a sufficiently short time,
then the success probabilities
for Game \ref{game:original} and Game \ref{game:tree} can only differ by a
small amount.

\begin{lemma}
For every algorithm $A$ for Game \ref{game:original} that makes at most
$t$ oracle queries, there exists an algorithm $A'$ for Game
\ref{game:tree} that also makes at most $t$ oracle queries such that for
all graphs $G$ and all colorings $C$,
\be
  \mathbb{P}_{\ref{game:original}}^{G,C} (A)
    \le \mathbb{P}_{\ref{game:tree}}^{G,C} (A') + O(t/2^{n})
\,.
\ee
\end{lemma}

\begin{proof}
Algorithm $A'$ simulates $A$, but whenever $A$ queries a name it has not
previously been sent by the oracle, $A'$ assumes the result of the query
is $11\ldots1$.  The chance that $A$ can discover the name of a vertex
that it is not told by the oracle is at most $t (2^{n+2}-2) / 2^{2n}$, and
unless this happens, the two algorithms will have similar behavior.
\end{proof}

Having restricted the algorithm to traverse a connected subgraph, we will
now show how to eliminate the coloring.  Consider the following game,
which is the same as Game \ref{game:tree} except that the oracle does not
provide a coloring:

\begin{game}\label{game:nocolors}
  The oracle contains a graph and a set of vertex names as described in
  Game \ref{game:original}.
  At each step, the algorithm sends the oracle the name of the $\ent$
  vertex or the name of a vertex it has previously been sent by the
  oracle.  The oracle then returns the names of the neighbors of that
  vertex.
  The algorithm wins it ever sends the oracle the name of the $\exit$
  vertex.
\end{game}

\noindent
For games that do not involve a coloring, we define the success
probability as
\be
  \mathbb{P}_{X}^G(A)
 = 
  \prob{\mathrm{names}}
       {\mbox{\rm algorithm $A$ wins game $X$ on graph $G$}}
\,.
\ee

\noindent
The next lemma shows that Game \ref{game:tree} and Game
\ref{game:nocolors} are equivalent:

\begin{lemma}
For any algorithm $A$ for Game \ref{game:tree} that makes $t$ queries to
the oracle, there is an algorithm $A'$ for Game \ref{game:nocolors} that
also makes at most $t$ queries to the oracle such that for all graphs $G$,
\be
  \expec{C}{\mathbb{P}_{\ref{game:tree}}^{G,C}(A)} 
          = \mathbb{P}_{\ref{game:nocolors}}^G(A') \,.
\ee
\end{lemma}

\noindent
Here $\expec{C}{\cdot}$ means the expectation is taken over the random
coloring of edges.

\begin{proof}
The only information needed to make up a coloring is whether a
vertex is in an even or odd column, and this information is always
available to an algorithm that can only traverse a connected subgraph.
Thus $A'$ can make up its own random coloring as it goes, and have the 
same probability of success as $A$.
\end{proof}

To obtain a bound on the success probability of Game \ref{game:nocolors},
we will compare it with a simpler game, which is the same except that it
provides an additional way to win:

\begin{game}\label{game:nocycles}
  The oracle contains a graph and a set of vertex names as described in
  Game \ref{game:original}.
  At each step, the algorithm and the oracle interact as in Game
  \ref{game:nocolors}.
  The algorithm wins it ever sends the oracle the name of the $\exit$
  vertex, or if the subgraph it has seen contains a cycle.
\end{game}

\noindent
Game \ref{game:nocycles} is clearly easier to win than Game
\ref{game:nocolors}, so we have

\begin{lemma}
For all algorithms $A$ for Game \ref{game:nocolors},
\be
      \mathbb{P}_{\ref{game:nocolors}}^{G}(A) 
  \le \mathbb{P}_{\ref{game:nocycles}}^{G}(A)
\,.
\ee
\end{lemma}

Now we further restrict the form of the subgraph that can be seen by the
algorithm unless it wins Game \ref{game:nocycles}.  
We will show that the subgraph an algorithm sees must be a
random embedding of a rooted binary tree.  For a rooted binary tree $T$,
we define an embedding of $T$ into $G$ to be a function $\pi$ from the
vertices of $T$ to the vertices of $G$ such that $\pi(\root) = \ent$ and
for all vertices $u$ and $v$ that are neighbors in $T$, $\pi(u)$ and
$\pi(v)$ are neighbors in $G$.  We say that an embedding of $T$ is {\em
proper} if $\pi(u) \not = \pi(v)$ for $u \ne v$.  We say that a tree $T$
{\em exits} under an embedding $\pi$ if $\pi(v) = \exit$ for some $v \in
T$.

We must specify what we mean by a random embedding of a tree.
Intuitively, a random embedding of a tree is obtained by setting
$\pi(\root) = \ent$ and then mapping the rest of $T$ into $G$ at random.
We define this formally for trees $T$ in which each internal vertex has
two children (it will not be necessary to consider others).  A random
embedding is obtained as follows:
\begin{enumerate}
\item Label the $\root$ of $T$ as 0, and label the other vertices of $T$
      with consecutive integers so that if vertex $i$ lies on the path
      from the root to vertex $j$, then $i$ is less than $j$.
\item Set $\pi (0) = \ent$.
\item Let $i$ and $j$ be the neighbors of $0$ in $T$.
\item Let $u$ and $v$ be the neighbors of $\ent$ in $G$.
\item With probability $1/2$ set $\pi (i) =u$ and $\pi (j) = v$, and
      with probability $1/2$ set $\pi (i) =v$ and $\pi (j) = u$.
\item For $i = 1, 2, 3, \ldots$, if vertex $i$ is not a leaf, and $\pi(i)$
      is not $\exit$ or $\ent$,
      \begin{enumerate}
	  \item Let $j$ and $k$ denote the children of vertex $i$, and let
                $l$ denote the parent of vertex $i$.
	  \item Let $u$ and $v$ be the neighbors of $\pi(i)$ in $G$ other
	        than $\pi(l)$.
	  \item With probability $1/2$ set $\pi(i)=u$ and $\pi(j)=v$, and
	        with probability $1/2$ set $\pi(i)=v$ and $\pi(j)=u$.
      \end{enumerate}
\end{enumerate}

We can now define the game of finding a tree $T$ for which a randomly
chosen $\pi$ is an improper embedding or $T$ exits:

\begin{game}\label{game:randomembedding}
  The algorithm outputs a rooted binary tree $T$ with $t$ vertices in
  which each internal vertex has two children.  A random $\pi$ is chosen.
  The algorithm wins if $\pi$ is an improper embedding of $T$ in $G_n'$ or
  if $T$ exits $G_n'$ under $\pi$.
\end{game}

\noindent
As the algorithm $A$ merely serves to produce a distribution on trees $T$,
we define
\be
  \mathbb{P}^G(T) =
  \prob{\pi}
       {\mbox{$\pi$ is an improper embedding of $T$ or $T$ exits $G$
        under $\pi$}},
\ee
and observe that for every distribution on graphs $G$ and all algorithms
taking at most $t$ steps,
\be
\max_{A} \expec{G}{\mathbb{P}^{G}_{\ref{game:randomembedding}}(A)}
   \leq  \max_{\mbox{\scriptsize trees $T$ with $t$ vertices}}
   \expec{G}{\mathbb{P}^{G} (T)}.
\ee

\noindent
Game \ref{game:nocycles} and Game \ref{game:randomembedding} are also
equivalent:

\begin{lemma}\label{lem:equivalence}
For any algorithm $A$ for Game \ref{game:nocycles} that uses at most $t$
queries of the oracle, there exists an algorithm $A'$ for Game
\ref{game:randomembedding} that outputs a tree of at most $t$ vertices
such that for all graphs $G$,
\be
    \mathbb{P}^{G}_{\ref{game:nocycles}}(A) 
  = \mathbb{P}^{G}_{\ref{game:randomembedding}}(A')
\,.
\ee
\end{lemma}

\begin{proof}
Algorithm $A$ halts if it ever finds a cycle, exits, or uses $t$ steps.
Algorithm $A'$ will generate a (random) tree by simulating $A$.  Suppose
that vertex $a$ in graph $G$ corresponds to vertex $a'$ in the tree that
$A'$ is generating.  If $A$ asks the oracle for the names of the neighbors
of $a$, $A'$ generates two unused names $b'$ and $c'$ at random and uses
them as the neighbors of $a'$.  Now $b'$ and $c'$ correspond to $b$ and
$c$, the neighbors of $a$ in $G$.  Using the tree generated by $A'$ in
Game \ref{game:randomembedding} has the same behavior as using $A$ in Game
\ref{game:nocycles}.
\end{proof}

Finally, we bound the probability that an algorithm wins Game
\ref{game:randomembedding}:

\begin{lemma}\label{lem:ourGraphs}
For rooted trees $T$ of at most $2^{n/6}$ vertices,
\be
 \max_{T}
  \expec{G}{\mathbb{P}^G(T)}
  \leq 3 \cdot 2^{-n/6}.
\ee
\end{lemma}

\begin{proof}
Let $T$ be a tree with $t$ vertices, $t \le 2^{n/6}$, with image $\pi(T)$
in $G_n'$ under the random embedding $\pi$.  The vertices of columns $n+1, n+2,
\ldots n+{n \over 2}$ in $G_n'$ divide naturally into $2^{n/2}$ complete
binary subtrees of height $n/2$.

\begin{enumerate}
\item[(i)]
It is very unlikely that $\pi(T)$ contains the root of any of these
subtrees, i.e., that $\pi(T)$ includes any vertex in column $n+{n \over
2}$.  Consider a path in $T$ from the root to a leaf.  The path has length
at most $t$, and there are at most $t$ such paths.  To reach column $n+{n
\over 2}$ from column $n+1$, $\pi$ must choose to move right ${n \over
2}-1$ times in a row, which has probability $2^{1-n/2}$.  Since there are
at most $t$ tries on each path of $T$ (from the {\sc root} to a leaf)
and there are at most $t$ such paths, the probability is bounded by
$t^2 \cdot 2^{1-n/2}$.

\item[(ii)]
If $\pi(T)$ contains a cycle, then there are two vertices $a,b$ in $T$
such that $\pi(a)=\pi(b)$.  Let $P$ be the path in $T$ from $a$ to $b$.
Then $\pi(P)$ is a cycle in $G_n'$.  Let $c$ be the vertex in $T$ closest
to the root on the path $\pi(P)$, and let $\pi(P)$ consist of the path
$\pi(P_1)$ from $c$ to $a$ and $\pi(P_2)$ from $c$ to $b$.

Let $S_1,S_2,\ldots,S_{2^{n/2}}$ denote the $2^{n/2}$ subtrees described
above.  Let $S_1',S_2',\ldots,S_{2^{n/2}}'$ denote the corresponding
subtrees made out of columns ${n \over 2}+1$ to $n$.  Without loss of
generality, let $\pi(c)$ be in the left tree of $G_n'$, i.e., in a column
$\le n$, as shown in Figure \ref{fig:subtrees}.

\begin{figure}
\setlength{\unitlength}{1.5pt}
\begin{picture}(150,220)
\drawline(65,0)(65,40)
\drawline(35,20)(65,40)
\drawline(35,20)(65,0)
\drawline(65,50)(65,90)
\drawline(35,70)(65,90)
\drawline(35,70)(65,50)
\drawline(65,120)(65,160)
\drawline(35,140)(65,160)
\drawline(35,140)(65,120)
\drawline(65,170)(65,210)
\drawline(35,190)(65,210)
\drawline(35,190)(65,170)
\drawline(85,0)(85,40)
\drawline(115,20)(85,40)
\drawline(115,20)(85,0)
\drawline(85,50)(85,90)
\drawline(115,70)(85,90)
\drawline(115,70)(85,50)
\drawline(85,120)(85,160)
\drawline(115,140)(85,160)
\drawline(115,140)(85,120)
\drawline(85,170)(85,210)
\drawline(115,190)(85,210)
\drawline(115,190)(85,170)
\put(50,98){\circle*{.75}}
\put(50,105){\circle*{.75}}
\put(50,112){\circle*{.75}}
\put(100,98){\circle*{.75}}
\put(100,105){\circle*{.75}}
\put(100,112){\circle*{.75}}
\put(15,130){\circle*{2}}
\put(57,135){\circle*{2}}
\dashline{2}(15,130)(55,190)
\dashline{2}(55,190)(95,140)
\dashline{2}(95,140)(57,135)
\dottedline{1}(57,135)(95,65)
\dottedline{1}(95,65)(45,20)
\dottedline{1}(45,20)(15,130)
\put(3,125){\makebox(10,10){\scriptsize $\pi(c)$}}
\put(35,115){\makebox(10,10){\scriptsize $\pi(a)\!=\!\pi(b)$}}
\put(47,125){\vector(1,1){7}}
\put(21,155){\makebox(10,10){\scriptsize $\pi(P_1)$}}
\put(14,80){\makebox(10,10){\scriptsize $\pi(P_2)$}}
\put(0,210){\makebox(10,10){$0$}}
\put(30,210){\makebox(10,10){$n \over 2$}}
\put(60,210){\makebox(10,10){$n$}}
\put(80,210){\makebox(10,10){$n+1$}}
\put(110,210){\makebox(10,10){$n+{n \over 2}$}}
\put(140,210){\makebox(10,10){$2n+1$}}
\end{picture}
\caption{Graphical representation of part (ii) of the proof of Lemma
\ref{lem:ourGraphs}.  The triangles represent the subtrees of $G_n'$ of
height $n/2$, the dashed line represents the path $\pi(P_1)$, and the
dotted line represents the path $\pi(P_2)$.  Together, these paths form a
cycle in the graph.}
\label{fig:subtrees}
\end{figure}
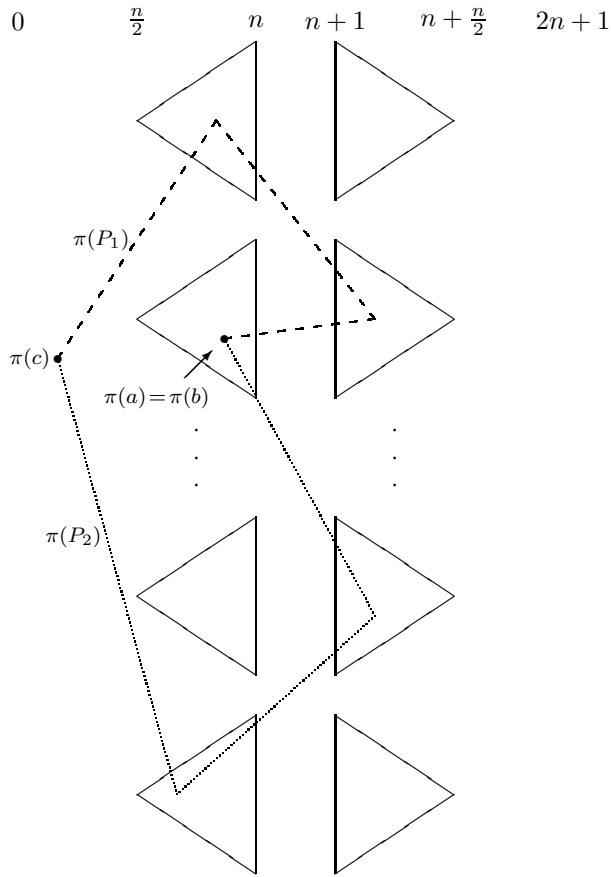

Now $\pi(P_1)$ visits a sequence of subtrees
$S_{i_1}',S_{j_1},S_{i_2}',\ldots$.  Similarly $\pi(P_2)$ visits a
sequence of subtrees $S_{k_1}',S_{l_1},S_{k_2}',\ldots$.  Since
$\pi(a)=\pi(b)$, the last subtree on these two lists must be the same.
(The other possibility is that $\pi(a)=\pi(b)$ does not lie in any
subtree, hence lies in columns $1$ through $n \over 2$ or $n+{n \over
2}+1$ through $2n$.  But the event that column $n+{n \over 2}$ is ever
reached has already been shown to be unlikely in part (i).  The same
argument bounds the probability of a return to column $n \over 2$ after a
visit to column $n+1$.)
At least one of the lists has more than one term (or all vertices visited
are in the left tree, which can't make a cycle).  The probability that the
last terms on the two lists agree is bounded by $2^{n/2}/(2^n-t)$, by the
construction of the random cycle that connected the two trees of $G_n'$.
As long as $t\le 2^{n-1}$, we have $2^{n/2}/(2^n-t)\le 2\cdot 2^{-n/2}$.
Since there are $t \choose 2$ paths $P$ to be considered, the probability
of a cycle is less than $t^2 \cdot 2^{-n/2}$.
\end{enumerate}
Overall we have shown that
\bea
  \expec{G}{\mathbb{P}^G(T)} 
    &\le& t^2 \cdot 2^{-n/2} + t^2 \cdot 2^{1-n/2} \\
    &\le& 3 \cdot 2^{-n/6}
\eea
if $t \le 2^{n/6}$.
\end{proof}

We have proved the following.

\begin{theorem}
Any classical algorithm that makes at most $2^{n/6}$ queries to the oracle
finds the $\exit$ with probability at most $4 \cdot 2^{-n/6}$.
\end{theorem}

%%%%%%%%%%%%%%%%%%%%%%%%%%%%%%%%%%%%%%%%%%%%%%%%%%%%%%%%%%%%%%%%%%%%%%%%%%%%%%%
\section{Discussion}\label{sec:discussion}

In this paper, we have applied a general implementation of (continuous
time) quantum walks to an oracular problem, and we have proved that this
problem can be solved efficiently by a quantum computer, whereas no
classical algorithm can do the same.  We considered the problem of finding
the name of the $\exit$ of a particular graph starting from the $\ent$,
given an oracle that outputs the names of the neighbors of an input
vertex.  Note that although our algorithm finds the name of the $\exit$,
it does not find a particular path from $\ent$ to $\exit$.

We can also view our problem as a decision problem by asking whether the
first bit of the name of the $\exit$ is $0$ or $1$.  Therefore, our result
can be also interpreted as showing the existence of a new kind of oracle
relative to which $\BQP \neq \BPP$ \cite{BV93,Sim94}.  We also note that,
although it is convenient to express our results in terms of a graph
traversal problem, the results can also be cast in terms of a graph
reachability problem, where one is given a graph and two vertices and the
goal is to determine whether there is a path connecting those vertices.

To efficiently simulate the quantum walk on a general graph, we required a
coloring of its edges.  In the specific case of a quantum walk on a
bipartite graph starting at the $\ent$, the walk can be constructed
without a coloring, as we describe in Appendix \ref{app:bipartite}.  It is
an open question whether one can find other techniques for implementing
quantum walks.  In particular, we do not know whether it is possible to
efficiently simulate a quantum walk on an arbitrary uncolored graph
without any restriction on the structure of the graph or the initial
state.

Many computational problems can be recast as determining some property of
a graph.  A natural question is whether there are other interesting
computational problems (especially non-oracular ones) that are classically
hard (or are believed to be classically hard) but that can be solved
efficiently on a quantum computer employing quantum walks.

%%%%%%%%%%%%%%%%%%%%%%%%%%%%%%%%%%%%%%%%%%%%%%%%%%%%%%%%%%%%%%%%%%%%%%%%%%%%%%%
\acknowledgments

We thank Jeffrey Goldstone for helpful discussions and encouragement
throughout this project. We also thank Dorit Aharonov for stimulating
correspondence about the implementation of quantum walks and John Watrous
for helpful remarks about quantum walks.

AMC received support from the Fannie and John Hertz Foundation, RC was
supported in part by Canada's NSERC, ED received support from the
Istituto Nazionale di Fisica Nucleare (Italy), and DAS was supported in
part by an Alfred P. Sloan Foundation Fellowship and NSF Award
CCR-0112487.
This work was supported in part by the Cambridge--MIT Foundation, by the
Department of Energy under cooperative research agreement
DE-FC02-94ER40818, and by the National Security Agency and Advanced
Research and Development Activity under Army Research Office contract
DAAD19-01-1-0656.

\appendix
%%%%%%%%%%%%%%%%%%%%%%%%%%%%%%%%%%%%%%%%%%%%%%%%%%%%%%%%%%%%%%%%%%%%%%%%%%%%%%%%
\section{An efficient classical algorithm to traverse the hypercube}
\label{app:hypercube}

In this appendix we describe a classical algorithm that traverses the
hypercube graph, where $\ent$ is any vertex and $\exit$ is defined to be
the unique vertex whose distance from $\ent$ is $n$.

The idea of the algorithm is to start at $\ent$ and repeatedly move one
level closer to $\exit$, where level $k$ is the set of vertices whose
minimum distance from $\ent$ is $k$.  In other words, the algorithm will
construct a sequence of vertices $a_0=\ent, a_1, \ldots, a_{n-1},
a_n=\exit$, where $a_k$ is in level $k$.  To choose the vertex $a_{k+1}$,
the algorithm will also require a set of vertices $S_{k-1}$, which is the
set of all neighbors of $a_k$ that are at level $k-1$.  Note that every
neighbor of $a_k$ is either at level $k-1$ or $k+1$.

The algorithm proceeds as follows.  First, let $a_1$ be any neighbor of
$a_0$ and let $S_0=\{a_0\}$.  Now suppose that (for some $k$) $a_k$ and
$S_{k-1}$ are known.  Then let $a_{k+1}$ be any neighbor of $a_k$ that is
not in $S_{k-1}$, and let $S_k$ contain every neighbor of $a_{k+1}$ that
is also a neighbor of some element of $S_{k-1}$.  The key point is that
the set $S_k$ constructed in this way is in fact the set of all neighbors
of $a_{k+1}$ whose distance from $a_0$ is $k$.  After $n$ such steps, the
algorithm reaches $a_n=\exit$.  Each step takes $O(n)$ queries, so the
total number of queries is $O(n^2)$.

%%%%%%%%%%%%%%%%%%%%%%%%%%%%%%%%%%%%%%%%%%%%%%%%%%%%%%%%%%%%%%%%%%%%%%%%%%%%%%%%
\section{Implementation of the quantum walk starting from a particular
vertex of a bipartite graph}
\label{app:bipartite}
 
In this appendix we outline an implementation of the quantum walk on a
bipartite graph with the initial state given by $\ent$ in the case where
the oracle does not provide an edge coloring. That is, the oracle is still
given by (\ref{eq:graphoracle}), but now $v_c(v_c(a))$ need not equal $a$.

This implementation uses the fact that, in any case, the neighbors of a
vertex must be given in {\it some} order, and this provides a coloring of
the directed edges. For a bipartite graph, this can be used to construct a
consistent coloring of the undirected graph.  For our particular graphs
$G_n'$, the construction is similar to the nine color construction found
in Section \ref{sec:lowerbound}.

The construction of the coloring requires a parity bit indicating which
side of the bipartite graph a vertex is in, and thus essentially requires
that the initial state is a particular vertex (e.g., the $\ent$) and not a
general superposition.  The $\ent$ is defined to be on side $0$ of the
bipartition, so the initial state has the parity bit equal to $0$.  By
flipping the value of the parity bit each time the oracle is queried, the
algorithm can keep track of the parity of each vertex.  In other words,
the name of each vertex can be assumed to contain a special bit that
indicates which side of the bipartition the vertex is on.

Using the parity bit, we can combine the colors of the directed edges to
form a consistent coloring of the undirected edges.  For vertices with
parity $0$, the color of each adjacent undirected edge is (incoming
color,~outgoing color), and for vertices with parity $1$, the color of
each adjacent undirected edge is (outgoing color,~incoming color).  The
color of an adjacent undirected edge can be easily computed using the
oracle and the parity bit, and it is also easy to compute whether there is
an undirected edge of a particular color incident on a given vertex.
Thus, the oracle that provides this consistent coloring of the undirected
edges can be simulated using the oracle that does not provide an edge
coloring.

%%%%%%%%%%%%%%%%%%%%%%%%%%%%%%%%%%%%%%%%%%%%%%%%%%%%%%%%%%%%%%%%%%%%%%%%%%%%%%%
% References

\end{document}